\documentclass[11pt]{article}

\usepackage{graphicx}
\usepackage{amssymb}
\usepackage{amsmath}
\usepackage{verbatim}
\usepackage{epsfig}
\usepackage{hyperref}

\newcommand{\beq}{\begin{equation}}
\newcommand{\eeq}{\end{equation}}
\newcommand{\beqa}{\begin{eqnarray}}
\newcommand{\eeqa}{\end{eqnarray}}

\newcommand{\tr}{{\rm Tr}}

\newcommand{\Lh}{\widehat\Lambda}
\newcommand{\Bt}{\bar B}
\newcommand{\nf}{{N_f}}
\newcommand{\nc}{{N}}
\newcommand{\A}{{\cal A}}
\newcommand{\xisb}{\xi_{{\rm SB}}}

\begin{document}

\begin{titlepage}
\begin{flushright}
{\large WIS/17/05-JUL-DPP\\
}
\end{flushright}

\vskip 1.2cm

\begin{center}

{\LARGE\bf Comments on the Meta-Stable Vacuum in $N_f=N_c$ SQCD and
  Direct Mediation}

\vskip 1.4cm

{\large  Andrey Katz$^a$, Yael Shadmi$^a$ and Tomer Volansky$^b$}
\\
\vskip 0.4cm
{\it $^a$Physics Department, \\Technion---Israel Institute of
  Technology, Haifa 32000, Israel}
\\
{\it $^b$Department of Particle Physics, 
\\Weizmann Institute of Science, Rehovot 76100, Israel}
\\

\vskip 4pt

\vskip 1.5cm

\begin{abstract}
\noindent  
We revisit $N_f=N_c$ SQCD and its non-supersymmetric minima
  conjectured by Intriligator, Seiberg and Shih (ISS).  We argue that
  the existence of such minima depends on the signs of three 
  non-calculable parameters and that no evidence can be inferred by
  deforming the theory.  We demonstrate this by studying a deformation
  of the theory which involves additional gauge singlets.  In this
  case, the conjectured minimum is destabilized.  We comment on the
  consequences of such singlets in models of direct mediation and in
  particular in the Pentagon model.
\end{abstract}

\end{center}

\vskip 1.0 cm

\end{titlepage}

\section{Introduction}
\label{sec:introduction}
The idea of dynamical supersymmetry breaking (DSB)~\cite{Witten:1981nf}
provides an elegant explanation of the hierarchy problem.  The
earliest examples were given in \cite{Affleck:1983mk,Affleck:1983vc}
and many more examples have been constructed since (for a review see
e.g. \cite{Shadmi:1999jy,Poppitz:1998vd}).  Still it has long been
understood that models which exhibit stable DSB vacua are non-generic.
Furthermore, while various tools for
constructing DSB models exist~\cite{Dine:1995ag}
there is no systematic classification of such theories.

DSB models are the starting point for generating
supersymmetry-breaking masses for the Standard Model (SM)
superpartners.  The breaking is communicated to the visible sector by
either gravity, or gauge interactions (for a review see
e.g.~\cite{Luty:2005sn, Shadmi:2006jv}) and therefore only soft
breaking is felt in the visible sector.  Most scenarios suffer from
various problems.  For example, in gravity-mediated models, flavor
changing contributions are not suppressed.  Minimal
anomaly-mediation~\cite{Giudice:1998xp, Randall:1998uk} leads to
tachyonic sleptons, and while this problem can be solved, the
solutions are typically fairly complicated.

Gauge mediated models, on the other hand, lead to
viable soft masses, with no ``hidden'' assumptions~\cite{Dine:1995ag,
Dine:1994vc, Giudice:1998bp}. 
Still, they are often deemed unattractive, since they involve
several tiers of messenger fields, including gauge singlets,
to mediate the breaking
from the the DSB model to the standard model.
Furthermore, the presence of the singlets often
leads to new supersymmetric color-breaking minima (although
the desired minimum is usually cosmologically 
stable)~\cite{Dasgupta:1996pz,Arkani-Hamed:1996xm}.
These aesthetic shortcomings have led people to seek models 
of ``direct gauge mediation'',
in which the standard model gauge group is embedded in
the unbroken global symmetry of the DSB 
model~\cite{Affleck:1984xz,Poppitz:1996fw,
Arkani-Hamed:1997fq,Arkani-Hamed:1997jv,Dimopoulos:1997ww, Murayama:1997pb}. 
While this avenue is indeed more compact, and does not generate
new unwanted minima, it typically results in Landau poles
below the Planck scale.
The reason is that, when the DSB model has a large
enough unbroken global symmetry to accommodate the standard model,
one finds too many new fields charged under the standard
model gauge group.

Recently, these ideas regained 
a lot of attention~\cite{Ooguri:2006pj,Banks:2006ma,
  Dine:2006gm,Murayama:2007fe,Dine:2006xt,Kitano:2006xg,Csaki:2006wi, 
Abel:2007uq},
following the elegant work of Intriligator, Seiberg and Shih
(ISS)~\cite{Intriligator:2006dd}.  As ISS show, by abandoning the
requirement of global supersymmetry breaking and allowing for 
meta-stable DSB vacua, one finds many more simple and generic
calculable models. 
In particular, ISS study
supersymmetric QCD (SQCD) with $N$ colors and $N < N_f < 3 N/2$
flavors, demonstrating that a meta-stable DSB vacuum is present near
the origin of field space.  The analysis strongly relies on
weak-strong Seiberg duality~\cite{Seiberg:1994pq} which 
provides a weakly coupled description
of the theory.

These constructions open new avenues for model building.  In
particular, near the origin of field-space the global symmetry is
large enough to allow for  embedding  the SM and therefore for
new models of direct mediation.  Nevertheless, such models still
suffer from Landau poles unless the supersymmetry breaking scale is pushed to
sufficiently large
scales~\cite{Kitano:2006xg, Csaki:2006wi, Intriligator:2006dd}. Such
models therefore turn out to be rather complicated.

ISS also consider the particularly interesting case of SQCD
with $N$ colors and $\nf=N$ flavors.
At low energy the theory is described by a non-linear sigma model with
a quantum-deformed moduli space.  Since these quantum corrections do
not allow all fields to be close to the origin of moduli space,
calculability is lost.  Thus there is no weakly coupled description of
the model that allows for establishing  a meta-stable DSB
vacuum.  Deforming the theory (to $N_f=N+1$) by adding another
flavor restores control, and by doing so ISS conjecture that a
meta-stable minimum exists also in the $N_f=N$ case.
 
The significance of this model lies in its minimal flavor symmetry.
Gauging this symmetry potentially does not introduce Landau poles at
low energy.  Thus this model is interesting for phenomenological
purposes.  Indeed, soon after the ISS discovery, the Pentagon
model~\cite{Banks:2006ma, Banks:2005df} was re-introduced,
demonstrating a simple and attractive realization of direct mediation.
Aside from the usual gauge dynamics, the model
contains a singlet which plays an important role in obtaining
a viable messenger spectrum, and which generates the $\mu$-term as well.

The importance of such models calls for further study of the
meta-stable minima in the $N_f=N$ case.  In this paper we reexamine
this case and its deformations.  As we discuss at some length, the
existence of a meta-stable minimum depends on the signs of three
non-calculable coefficients which appear in the K\"ahler potential.
In the deformation that ISS consider, these parameters are irrelevant
by construction, so that the non-SUSY minimum is calculable.  We
suggest another deformation which is closely related to the
Intriligator-Thomas-Izawa-Yanagida (ITIY) DSB
model~\cite{Intriligator:1996pu, Izawa:1996pk}.  As in the $N_f=N+1$
deformation, there exists a region of parameter space where the theory
is calculable and the ISS-like extremum (which coincides with the
conjectured minimum as we approach $\nf=N$ SQCD) is found to be a
saddle point rather than a minimum.  This demonstrates the weakness of
the conjecture, implying that the $N_f=N$ extremum is just as likely
to be a saddle-point.

One approach towards settling this issue is to take advantage of the
AdS/CFT correspondence.  In~\cite{Argurio:2006ny} the $N_f=N$ model
was realized on fractional branes placed on a ${Z}_2$ orbifold of the
conifold.  A gravity dual was suggested and found to posses a
non-supersymmetric state, indicating that the conjectured meta-stable
minimum indeed exists.  While clearly a step in the right direction, a
complete gravity solution is still missing and more importantly, it is
not clear whether such non-supersymmetric states remain in the
transition between large 't~Hooft and weak gauge coupling.  We believe
further work is needed in this regard.

The appearance of a saddle-point is directly related to the
introduction of new gauge-singlet degrees of freedom.  Large couplings
to the singlets may destabilize the desired minimum.  It is therefore
natural to question the validity of direct-mediation models which take
advantage of the $N_f=N$ scenario, and in particular of the Pentagon
model~\cite{Banks:2005df,Banks:2006ma}.  Unlike the deformation
discussed above, this model is non-calculable so one cannot reliably
establish the existence of a non-susy minimum with a viable messenger
spectrum.  Still, we argue that for small quark masses, such a minimum
requires a large meson-singlet coupling, which would probably
destabilize the minimum.  Moreover, because of the large coupling,
even if a minimum exists, it is not directly related to the ISS
conjectured minimum.

The paper is organized as follows.  In
section~\ref{sec:iss-conjecture} we review the ISS $N_f=N$ conjecture,
emphasizing the $N_f=N+1$ deformation and its relation to the original
theory.  In section~\ref{singlets} we consider a different
deformation, in which the mesons and baryons are coupled to singlet
fields.  We show that the ISS-like extremum is in fact a saddle-point
in the region of parameter space where the model is calculable.  In
Section~\ref{sec:direct-medi-with} we consider direct mediation in the
Pentagon model.  Some details of the calculation are given in the
Appendix.

\section{The ISS conjecture}
\label{sec:iss-conjecture}
We begin this section with a quick review of the 
ISS supersymmetry-breaking minima for $SU(N)$ SQCD with 
$\nc+1<\nf\le 3\nc/2$~\cite{Intriligator:2006dd}.
For this range of $\nf$, the IR theory can be described by
the weakly coupled ``magnetic'' theory, with $\nf-\nc$ colors
and with the superpotential
\begin{eqnarray}
  \label{eq:45}
  W = \tr m_Q M + \frac{1}{\Lh}\tr qM\bar q \ . 
\end{eqnarray}
Here $M$ corresponds to the meson of the
original, ``electric'' theory, $q$, $\bar q$, are the magnetic quarks,
and $m_Q$ is the (electric) quark mass. 
The scale $\Lh$ is related to the strong coupling scales of the
electric and magnetic theories. 
The superpotential also contains non-renormalizable terms generated by
non-perturbative effects.
These are essential for seeing the supersymmetric minima, 
but are negligible close to the origin.
The potential is minimized at
\begin{eqnarray}
  \label{eq:96}
  M=0, \quad q=-\bar q = \left(\begin{array}{c}
      q_0\\0
    \end{array}\right), \qquad q_0^2 = m_Q\,\Lh\,\mathbf{1}_{N_f-\nc}\,,
\end{eqnarray}
where $M$ is an $N_f\times N_f$ matrix and $q$, $\bar q$ are
$(N_f-\nc)\times N_f$.  At the minimum, the $F$-terms for some $M$'s
are nonzero and supersymmetry is broken.  Some of the dual quarks and
mesons get mass at tree-level, through the cubic term of the
superpotential~\eqref{eq:45}.  This cubic interaction also generates
masses at the loop level for the remaining massless scalars 
(apart from the Goldstone bosons).
For small
$m_Q$, these masses are parametrically larger than contributions from
non-calculable corrections to the K\"ahler potential.  Thus, the
theory near the origin is calculable by virtue of two important
ingredients: (i) the smallness of $m_Q$, and (ii) the cubic
superpotential interaction, which generates positive masses for all
the scalar fields.  This second ingredient is missing for $\nf=\nc$.

Let us consider then $N_f=\nc$.  At low energy, the theory is
described by a non-linear sigma model with the 
superpotential~\cite{Seiberg:1994bz}
\begin{eqnarray}
  \label{eq:104}
  W = m_Q\tr M + {\cal A}(\det M - B\Bt - \Lambda^{2\nc})\,.
\end{eqnarray}
Here for simplicity, we take the quark masses to be 
${m_Q}_{ij} = m_Q\delta_{ij}$.  The non-dynamical auxiliary field
${\cal A}$ is introduced to enforce the quantum constraint.  The
theory has $\nc$ supersymmetric minima at
\begin{eqnarray}
  \label{eq:121}
  M_{ij} = \Lambda^2\frac{(\det m_Q)^{1/\nc}}{m_Q}\;\delta_{ij}, \qquad
  B=\Bt = 0.
\end{eqnarray}
As a first attempt at finding a non-supersymmetric minimum we
extremize the potential on the baryonic branch, assuming a canonical
K\"ahler potential.  One finds a classical moduli space of solutions
with $\A=0$ and non-vanishing baryon number.  We concentrate on the
point with the largest flavor symmetry,
\begin{eqnarray}
  \label{extr}
  M=0, \qquad B=- \Bt = \Lambda^{\nc}.
\end{eqnarray}
The discussion below can be carried over to any other extrema.
Around~\eqref{extr}, only $B_-\equiv(B- \Bt)/\sqrt{2}$ is massive due
to the quantum constraint.  The combination $B_+\equiv (B+
\Bt)/\sqrt{2}$ \footnote{Strictly speaking, the dynamical field is not
  $B_+$ but rather $b$ where $B=\Lambda^Ne^{b}$, $B=-\Lambda^Ne^{-b}$
  \cite{Intriligator:2006dd}.  Still we will find it more convenient
  to work with $B_+$. The two parametrizations coincide to quadratic
  order.}, as well as all the mesons, remain massless: there is no
superpotential coupling that can generate masses for these fields.  In
order to discover the nature of this extremum, one must therefore take
into account corrections to the K\"ahler potential.  The only non-zero
$F$-term is the meson $F$-term, $F_M$, so the relevant quantity is the
$M-M^\dagger$ entry of the inverse K\"ahler metric. 
The K\"ahler potential is of the form
\begin{eqnarray}
 \label{kahler}
 K = \frac{\tr M^\dagger M}{\Lambda^2} + 
\frac{(B_+ + B_+^\dagger)^2}{\Lambda^{2N-2}} +
c_1 \frac{\tr M^\dagger M M^\dagger M}{\Lambda^4} +
c_2  \frac{{(\tr M^\dagger M)}^2}{\Lambda^4} +
c_3\frac{{(B_+ + B_+^\dagger)}^2 \tr{M^\dagger M}}{\Lambda^{2N+2}}
+ \cdots . 
\end{eqnarray}
Here $c_1$, $c_2$ and $c_3$ are order-one parameters\footnote{For
  simplicity, we ignore order-one coefficients in front of the leading
  terms in the K\"ahler potential.  Thus one should not confuse the
  above parametrization with the one of~\cite{Intriligator:2006dd}.},
and the ellipses stand for terms which are irrelevant for our
discussion.  The form of~(\ref{kahler}) follows from the flavor,
baryon and non-anomalous ${\mathbb Z}_{2N}$ axial symmetries.  The
last three terms in \eqref{kahler} are small for $M/\Lambda^2,
B_+/\Lambda^{N} \ll 1$.  However, they are the only source of meson
and $B_+$ masses.  Indeed, the potential takes the form,
\begin{eqnarray} 
  \label{corrections}
  V \sim \left(1+\alpha\frac{\tr M^\dagger M}{\Lambda^4} 
+ \tilde\beta \frac{\tr M \tr M^\dagger}{\Lambda^4} + 
\gamma \frac{(B_+ + B_+^\dagger)^2}{\Lambda^{2N}}\right)
\,\left|m_Q \Lambda\right|^2 + ...,
\end{eqnarray} 
where the coefficients $\alpha$, $\tilde\beta$, and $\gamma$
depend on $c_1$, $c_2$ and $c_3$.
Therefore the above corrections contribute order $|m_Q|^2 \ll
\Lambda^2$ to the masses-squared of the canonically-normalized 
meson and baryon.  
In order for this extremum to be a minimum, we must have
\begin{eqnarray}
  \label{eq:116}
\alpha>0\,, \ \ \ \beta \equiv \alpha/N+\tilde\beta>0\,,\  \ \ \gamma > 0\,.
\end{eqnarray}
However, as ISS discuss, the theory is strongly coupled at the scale 
$\Lambda$, and so  $\alpha$, $\beta$ and $\gamma$ are non-calculable.   
At the field theory level one can therefore only
conjecture the existence of a supersymmetry-breaking minimum near the origin.
Moreover, studying other minima far away from the origin
at $M/\Lambda \sim 1$ requires the knowledge of higher order terms 
in the K\"ahler potential.

To make further progress, ISS deformed the theory by adding another
flavor.  This is the $N_f=\nc+1$ case which we now discuss.  
To understand this deformation, one must carefully follow
the corrections to the K\"ahler potential, as the mass of the
extra flavor is dialed.  
To be concrete, let us take 
${m_Q}_{ij} = {\rm diag}(m_Q,..,m_Q,m_{N+1})$ with 
$m_{N+1} \geq m_Q$.
At low energy, this theory too is described by a non-linear sigma
model in terms of the baryons and mesons~\cite{Seiberg:1994bz}. 
The superpotential is
identical to the superpotential of the $\nf>\nc+1$ magnetic theory
with the dual quarks replaced by the baryons~\cite{Seiberg:1994bz}, 
\begin{eqnarray}
  \label{eq:118}
  W = \frac{1}{\widehat\Lambda^{2\nc-1}}
\left(\widehat B\widehat M\widehat {\bar B} - \det \widehat M\right) 
+ \tr m_Q \widehat M.
\end{eqnarray}
Here $\widehat B$ and $\widehat{\bar B}$ are the $\nc+1$ baryons,
$\widehat M$ are the mesons of the deformed theory and $\Lh$ is the
scale at which the theory becomes strongly coupled.  Unlike in the
discussion of the magnetic theory at the beginning of this section,
here we chose to display the non-renormalizable term $\det \widehat M$
since it is important for recovering the $\nf=\nc$ superpotential.
Still, as before, this term will play no role in the analysis near the
origin.

To make contact with the $\nf=\nc$ theory, 
it is convenient to write the $\nf=\nc+1$ (hatted) fields as 
\begin{eqnarray}
  \label{eq:119}
  \widehat M &=& \left(\begin{array}{cc}
      M_i^j & \widehat M_i^{N+1}\\
      \widehat M_{N+1}^j & M_{N+1}^{N+1}
    \end{array}\right), \qquad
\\
\widehat B &=& (B^i, B) \qquad 
\widehat{\bar B} = (\bar B^i, \bar B)\ .
\end{eqnarray}
As $m_{N+1}\rightarrow \infty$, the heavy flavor can be integrated
out, leaving only $M$, $B$ and $\bar B$ light.  In this limit, the
theory reduces to the original $N_f=\nc$ case with the identification
$\A = M_{N+1}^{N+1}/\widehat\Lambda^{2\nc-1}$ and 
$\Lambda^{2\nc} = m_{N+1}\widehat\Lambda^{2\nc-1}$.  
Indeed, for constant
$\Lambda$, the limit $m_{\nc+1}\rightarrow \infty$ corresponds to
$\widehat\Lambda \rightarrow 0$ which sets the kinetic term of $\A$ to
zero making it non-dynamical.

For finite $m_{N+1}$ however, $M_{N+1}^{N+1}$ must be treated 
as a dynamical
field.  As before, one may try to minimize the tree-level 
potential first, ignoring corrections to the K\"ahler potential.  
The analysis is identical to the the analysis of the $\nf>\nc+1$ theory,
only now one quark mass is different.
Again, we concentrate on the extremum, 
\begin{eqnarray}
  \label{eq:120}
  \widehat M=0,\qquad B^i=\bar B_i = 0, 
\qquad B=-\bar B = \Lambda^N,
\end{eqnarray}
just as for $\nf=\nc$.  But as opposed to the $N_f=\nc$ theory, the
superpotential~\eqref{eq:118} contains a cubic term.  At tree level,
this term generates a mass-squared of order $m_{N+1}\widehat\Lambda$
for all fields apart from $M$ and $B_+\equiv (B+\bar B)/\sqrt{2}$.
The latter, just as for the case of more flavors, become massive at
the one loop level, with masses of order
$m^2_Q\widehat\Lambda/m_{N+1}$.  To see this, note that the only
fields with non-zero $F$-terms are $M_i^i$ ($i\leq\nc$), with $F\sim
{m_Q} \widehat\Lambda$.  As a result, the fields $B_i$ and $\bar B^i$
have supersymmetric masses-squared of order $m_{N+1}\Lh$, and
supersymmetry-breaking masses-squared of order $m_Q\Lh$.  These fields
then generate a non-zero supertrace, leading to masses for $M$ and
$B_+$, \beq\label{loopmass} m^2_{{\rm loop}}\sim \frac1{16\pi^2}\,
\frac{m_Q^2\Lh}{m_{N+1}}\ .  \eeq This is the crucial difference
between the original $N_f=\nc$ model and the deformation: in the
deformed theory, just as for larger values of $\nf$, all scalars apart
from the Goldstones get masses either at tree-level or at one-loop,
and the pseudo-flat directions are (at least naively) lifted, giving a
minimum at~\eqref{eq:120}.

However, on top of these mass terms, one must still consider the
corrections to the K\"ahler potential.  As in
eqn.~\eqref{corrections}, these contribute $\delta m^2\sim m_Q^2$ and
are therefore negligible compared with~\eqref{loopmass} {\it as long
  as $m_{\nc+1} \ll \widehat \Lambda$}.  Thus for sufficiently small
$m_{\nc+1}$, we can reliably establish a true minimum.  On the other
hand, for $m_{\nc+1} \geq\widehat \Lambda$, the
signs of the
coefficients $\alpha$, $\beta$ and $\gamma$  of 
eqns.~\eqref{corrections},~\eqref{eq:116} 
(with $\Lambda$ replaced by $\Lh$) are crucial.

Let us therefore summarize the essence of the ISS conjecture.  Whether
the point~\eqref{extr} is a minimum or not depends on the signs of
unknown parameters, $\alpha$, $\beta$ and $\gamma$. One can
deform the theory by adding tree-level couplings which stabilize the
above extremum by generating positive masses-squared for all fields.
The deformation can be worked out in a limit where the above
parameters are not important and can be neglected.

Clearly, the deformation gives us no information on $\alpha$, $\beta$
and $\gamma$.  It is therefore just as likely that one or more of the
mesons and baryons is tachyonic.  Physically this would amount to a
smooth transition in the potential as the minimum becomes a
saddle-point when $m_{\nc+1}$ crosses $\Lh$ from below.  To emphasize
this point, we now consider a different deformation of the $\nf=\nc$
theory, with the mesons and baryons coupled to singlet fields.  As we
will see, when the deformed theory is calculable, the
extremum~\eqref{extr} turns out to be a saddle point, demonstrating
that such a transition indeed occurs.  We therefore conclude that no
information can be extracted on the nature of the $N_f=\nc$
supersymmetry-breaking extremum by deforming the theory.

\section{Adding singlets} 
\label{singlets}

We  now deform the ISS model by adding singlet fields 
$S_{ij}$, $T$ and $\bar{T}$
with superpotential couplings to the mesons and baryons,
\beq 
 \label{superpot}
  W= m_Q\tr M + \lambda \tr S M +\kappa (T B + \bar{T} \bar{B}) 
  + \frac{1}{2}m_S \tr S^2+ \frac{1}{2}m_T(T^2+\bar{T}^2)\,.
\eeq
This is nothing but the Intriligator-Thomas-Izawa-Yanagida model
(ITIY)~\cite{Intriligator:1996pu, Izawa:1996pk}, with singlet mass
terms added.  Without these mass terms, the quark masses can be
absorbed by a shift redefinition of the singlets
$S_{ij}$.

As we will see below, the model has a local non-supersymmetric
extremum similar to the minimum conjectured by ISS.  As the singlets
decouple, the model approaches $\nf=\nc$ SQCD, and the local
supersymmetry-breaking extremum approaches the ISS-conjectured
minimum~(\ref{extr}).  We can decouple the singlets either by
decreasing their superpotential couplings to the mesons and baryons,
or by increasing their masses,
\begin{eqnarray}
  \label{decoupling}
  \lambda \to 0 \ \ {\rm or}\quad m_S\to \infty\ ;\ \ \ 
   \kappa\to 0 \ \ {\rm or}\quad m_T\to \infty\,.  
\end{eqnarray}
However, as we will see below, there is a lower bound on the
couplings $\lambda$, $\kappa$, and equivalently, an upper
bound on the masses $m_T$, $m_S$. 
For very small couplings (or very large masses), non-calculable 
K\"ahler corrections become important and we cannot reliably
study the ISS-like extremum, much like in the $\nf=\nc+1$ deformation.
Still, as long as the model is calculable, we will find that
this extremum is a saddle point rather than a minimum.

\subsection{Supersymmetric minima}
Before going on, it is useful to recall what happens in the ITIY
model.  The classical superpotential of the model is given
by~(\ref{superpot}) with $m_S$ and $m_T$ set to zero.  Supersymmetry
is then broken, since the singlet $F$-terms only vanish when the
mesons and baryons are at the origin, in conflict with the
quantum-modified constraint.  Defining again
$T_\pm=(T\pm\bar{T})/\sqrt2$, one finds that at tree-level, $T_-$ is a
flat direction.  This degeneracy is lifted at the loop-level, and as
argued in~\cite{Chacko:1998si}, the loop corrections can be reliably
computed near the origin.  Indeed these loop corrections are generated by
light states, and scale as ${\cal O}(\kappa^4)$, while non-calculable
corrections from states at the scale $\Lambda$ are suppressed by
${\cal O}(\kappa^6)$~\cite{Chacko:1998si}.  Thus for sufficiently
small $\kappa$, all fields are stabilized at the origin, apart from
$B_-=\sqrt{2}\Lambda^N$.  Since the only nonzero $F$-term is $F_{T_-}$, the
Goldstino is the $T_-$ fermion.

As discussed above, here we add singlet mass terms.
As these masses are turned on, supersymmetric 
vacua move in from infinity, and the theory can only
have {\it local} supersymmetry-breaking minima at best.
Taking into account  the quantum-modified constraint,
\beq
W_{NP}={\cal A}
\left(\det M 
    -B \bar{B} - \Lambda^{2N}\right) \,,
\eeq
one finds three families of supersymmetric solutions.
The first is given by 
\beq
  \label{susysol1}
  B_\pm=T_\pm=0\,,\qquad |M|= \Lambda^2\,,\qquad |S| =
  -\frac{\lambda}{m_S}\Lambda^2\,.
\eeq
The second is given by, up to terms of order $\lambda^2$,
\begin{eqnarray}
\label{susysol2}  
&&B_+=T_+=0\,,\qquad M \simeq
  -\left(\frac{m_Q m_T}{\kappa^2}\right)^{1/(N-1)}\,,\qquad S \simeq
  \frac{\lambda}{m_S}\left(\frac{m_Q
      m_T}{\kappa^2}\right)^{1/(N-1)}\,,
\nonumber \\
  &&B_-^2 = -\det M + \Lambda^{2N}\,,\qquad T_- = -\frac{\kappa}{m_T}
  B_-\,,
\end{eqnarray}
and the third solution is obtained from the second for 
$B_+^2\leftrightarrow -B_-^2$, $T_+\leftrightarrow T_-$,
and $M \rightarrow -M$, $S \rightarrow -S$.
Clearly, in the decoupling limit~(\ref{decoupling}),
only the first solution remains at a finite distance from
the origin. The other two solutions approach the classical
solutions with the meson VEVs running to infinity.

\subsection{Non-supersymmetric saddle points}
\label{sec:non-supersymm-saddle}
We are now ready to look for the ISS conjectured minimum.  For now, we
will assume that the K\"ahler potential is canonical in all fields.
We will later examine the region of validity of this approximation.
Strictly speaking, around a given solution we should use the
constraint to eliminate the heavy degree of freedom, say $B_-$, and
derive the potential for the remaining degrees of freedom.  In the
process, various non-renormalizable interactions of the remaining
fields will be induced, making the potential quite unwieldy.  We will
therefore first perform the analysis with the Lagrange multiplier in
place, and later explain how the results are modified in the full
analysis (this analysis is described in detail in the Appendix).

It is simple to verify that at the minimum, the $F$-terms of $B_\pm$,
$T_\pm$ and $\A$ all vanish, resulting in two possible solutions
at\footnote{There is another uninteresting solution with
  $B_+=B_-=T_+=T_-=0$.  This solution is on the mesonic branch and is
  not related to the ISS conjecture.}, 
\begin{eqnarray}
  \label{ourextr}
  B_\pm=T_\pm=0\,,\qquad \A =\pm\frac{\kappa^2}{m_T}\,,\qquad  
B_\mp^2 = \mp 2(\det M - \Lambda^{2N})\,,\qquad T_\mp = -\frac{\kappa}{m_T}
  B_\mp\ .
\end{eqnarray}
Since we are interested in extrema which preserve the $SU(\nc)_{{\rm
    diag}}$ global symmetry, we take the ansatz,
\begin{eqnarray}
  \label{diagansatz}
  M_{ij} = M \,\delta_{ij} \,,\qquad S_{ij} = S\,\delta_{ij}\ .
\end{eqnarray}
One therefore obtains an effective potential for $M$ and $S$,
\begin{eqnarray}
  \label{eq:15}
   V=N\left| \lambda S +m_Q\pm\frac{\kappa^2}{m_T}M^{N-1}\right| ^2
+N\left| \lambda
  M+mS\right| ^2\ .
\end{eqnarray}
with a non-supersymmetric extremum  at,
\begin{eqnarray}
  \label{ourextrms}
  M&=&  
\left(\pm\frac{\lambda^2}{(\nc-1) m_S} \,
       \frac{m_T}{\kappa^2}\right)^{\frac{1}{\nc-2}}\,,
  \\
   S&=& -\frac{\lambda^*}{|\lambda\Lambda|^2+|m_S|^2}\left[m_Q\Lambda^2 +
\left(\pm\frac{\lambda^2}{(\nc-1) m_S} \,
         \frac{m_T}{\kappa^2}\right)^{\frac{1}{\nc-2}}\,
\left(\frac{\lambda}{\lambda^*}m_S^*+ \frac{\lambda^2\Lambda^2}{(\nc-1) m_S}\right)\right]\,.\nonumber
\end{eqnarray}

We now wish
to relate the above solutions to the ISS extremum.
We therefore consider the case $\A>0$.  To this end, one may take the
decoupling limit, \eqref{decoupling}, in various ways thereby probing
the space of vacua in the original $N_f=N$ SQCD.  To discover the
nature of the extrema, one then needs to compute the mass
spectrum for each given decoupling limit.  In particular, to
approach the ISS extremum, the singlets $S$ should decouple faster
than $T_\pm$, for example by taking
\begin{eqnarray}
  \label{lambdakappa}
  \lambda, \kappa \rightarrow 0\,,\qquad 
\frac{\lambda}{\kappa}\rightarrow 0\,.
\end{eqnarray}
The only nonzero $F$-terms at this extremum are $F_M$ and $F_S$,
with $F_M\sim m_Q$ for small $\lambda$.
Thus the Goldstino is a mixture of the $M$ and $S$ fermions,
and it tends to the mesino when $\lambda\to 0$, as expected
for the ISS minimum.   

We can now analyze the nature of this extremum.  
To leading order in $\lambda$, the $(S,M)$
mass-squared matrix takes the simple form,
\begin{eqnarray}
  \label{massmatrix}
  m_{{\rm bosons}}^2 = N\left(\begin{array}{cccc}
      |\lambda \Lambda|^2 & \lambda^* \Lambda m_S &  \xisb &0 \\
      \lambda m_S^* \Lambda & \vert m_S\vert^2 & 0 &0\\
      \xisb^* & 0 & |\lambda \Lambda|^2 & \lambda \Lambda m_S^*\\
      0 & 0 & \lambda^* m_S \Lambda & |m_S|^2 \\
    \end{array}\right) \,,
\end{eqnarray}
where
\begin{eqnarray}
  \label{eq:1}
  \xisb \simeq (N-1)(N-2)\frac{\kappa^2}{m_T}
  \left(\frac{\lambda^2}{(N-1)m_S}
    \frac{m_T}{\kappa^2}\right)^\frac{N-3}{N-2}\, 
  m_Q^* \, \Lambda^4\ ,
\end{eqnarray}
is the supersymmetry-breaking contribution.  The determinant of the
matrix above is negative, so there is at least one tachyonic
direction. In fact it is easy to see that there is precisely one such
direction.  Note that the determinant of the diagonal $2\times 2$
block, which coincides with the fermion mass matrix, is exactly zero,
signalling the presence of the Goldstino.  The $S$ and $M$ fermions
mix to give one massive state, which is predominantly $S$ of mass near
$m_S$, and one massless fermion, the Goldstino, which is mostly $M$.
The supersymmetry-breaking contribution $\xisb$ results in splittings
between the fermions and scalars.  For small $\lambda$, this splitting
occurs mostly in the $M$ sector.  Since however the supertrace still
vanishes (we are working at tree-level) one scalar becomes lighter
than the Goldstino, with a tachyonic mass $m^2\sim -|\xisb|$.

\begin{center}
\begin{figure}\label{fg}
 \begin{center}
 \includegraphics[width=80mm]{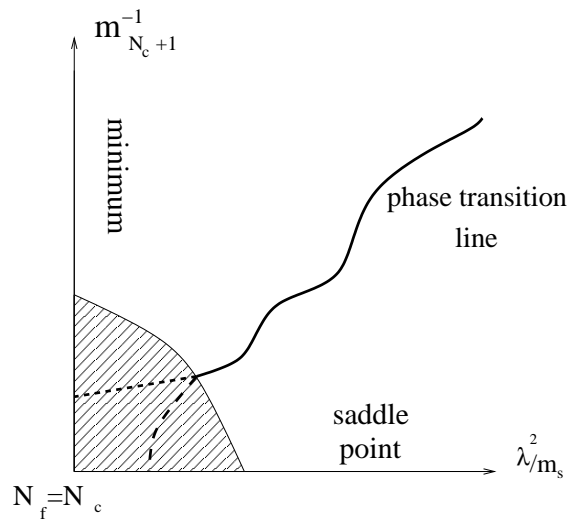}
\end{center}
\caption{The two deformations of $\nf=\nc$ SQCD. 
The transition between a saddle point and a minimum occurs in the
shaded region where calculability is lost.}
\end{figure}
\end{center}

As we saw in section \ref{sec:iss-conjecture}, the meson masses also receive 
contributions from non-calculable K\"ahler terms,
which are of order $\vert F_M\vert^2/\Lambda^2 \sim m_Q^2$
[see eqn.~(\ref{corrections})]\footnote{
In fact, due to the new interactions with the singlets, there are additional
non-calculable contributions to the K\"ahler potential which are
negligible for small $\lambda$, $\kappa$.}. 
For the model to be calculable, these contributions must be smaller
than the smallest eigenvalue of the mass-matrix~(\ref{massmatrix}),
\begin{eqnarray}
  \label{eq:17}
  \xisb \gg m_Q^2 \  . 
\end{eqnarray}
One can choose, for example (for large $N$), 
\begin{eqnarray}
  \label{eq:19}
  m_Q\ll \lambda \Lambda \ll m_S \lesssim \Lambda\ .  
\end{eqnarray}
A similar bound holds in the baryonic sector for $\kappa^2/m_T$.  Thus
we see that we cannot decouple the singlets completely while
preserving the calculability of the model.  There is a lower bound on
the coupling $\lambda$, or, alternatively an upper bound on the mass
$m_S$.  Outside the allowed range, the 
signs of 
the parameters $\alpha$, $\beta$ 
and $\gamma$  become crucial for establishing  a minimum.  In
this regard the above deformation is on exactly the same footing as
the deformation considered by ISS. 
 This
 situation is depicted in figure~\ref{fg}. 

As we noted above, so far we worked with the Lagrange multiplier for
simplicity. In the Appendix we present a more careful analysis,
where we eliminate one of the fields using the constraint from the
start.  Indeed, the location of the
extremum~(\ref{ourextr}),(\ref{ourextrms}), and the mass-squared
matrix~(\ref{massmatrix}) are corrected by small amounts, but the
conclusion remains unchanged.

One could still hope that the instability we found might be cured by
positive contributions arising from the Coleman-Weinberg (CW)
potential.  The situation is different from the $N_f\ge \nc+1$ case.
As we reviewed in the previous section, for $\nf=\nc+1$ there are
pseudo-flat directions which are lifted by the dominant, calculable
one-loop corrections.  In our case on the other hand, there are no
such flat directions at tree level. Furthermore, the one-loop
contributions are smaller than the tree-level ones.  Note that the
only contribution to the CW potential is from the fields $M$ and $S$,
since the masses of the remaining fields are approximately
supersymmetric.  Thus the CW potential is roughly
\begin{eqnarray}
  \label{CW}
  \Delta V=\frac{1}{64\pi^2}{\rm Str} M^4\log
  \frac{M^2}{\Lambda^2} \propto
  \frac1{64\pi^2}\,  \xisb^2~.
\end{eqnarray}
This correction is one loop-suppressed compared with the tree-level
contribution to the tachyonic mass and thus cannot stabilize the
extremum.

Finally, one could ask whether the CW potential can generate a
distinct minimum which coincides with the ISS minimum in the
decoupling limit.  While a minimum is indeed generated for
sufficiently small singlet mass $m_S$, in the decoupling limit this
minimum is infinitely far in field space from the ISS and
supersymmetric minima.  The simplest way to see this, is to
consider the one loop correction to the K\"ahler potential for $S$,
\begin{eqnarray}
  \label{eq:23}
  \delta K \sim -\frac{N}{32\pi^2} |\lambda S+m_Q|^2
  \log\frac{|\lambda S+m_Q|^2}{\Lambda^2}\ .
\end{eqnarray}
Therefore for sufficiently small mass, $m_S$, a local minimum is
generated at $S \sim -m_Q/\lambda$. 
In the decoupling limit this
minimum is driven to infinity. Hence, it cannot correspond to the
ISS conjectured minimum.

\section{Direct mediation with singlets?} 
\label{sec:direct-medi-with}
As discussed in the introduction, one of the main virtues of the ISS
supersymmetry-breaking minima is that many fields are at the origin.
There is thus a large unbroken global symmetry, which makes these
theories promising starting points for models of direct gauge
mediation~\cite{Affleck:1984xz,Poppitz:1996fw,
Arkani-Hamed:1997fq,Arkani-Hamed:1997jv,Dimopoulos:1997ww, Murayama:1997pb}.  
In particular, the most compact model which
potentially does not lead to Landau poles for the standard model
couplings at low energy is the $N_f=N$ case.  Recently such a model
has been proposed, taking advantage of the conjectured minimum in
$N_f=\nc$ SQCD~\cite{Banks:2006ma}\footnote{In fact, an earlier 
version of the model~\cite{Banks:2005df}, which is not based
on the ISS minimum, involves a meta-stable supersymmetry-breaking
minimum in the context of 
``Cosmological Supersymmetry Breaking''~\cite{Banks:2000fe}.}. 
The model is based on the specific case of $N_f=\nc=5$ 
(and hence dubbed the ``Pentagon model''),
with the SM gauge group embedded in the $SU(5)_{\rm
  diag}$ global symmetry.  
One gauge singlet, $S$, is added to the
model, in order to generate the $\mu$-term through the superpotential
coupling $SH_uH_d$.  $S$ obtains a VEV of order the
supersymmetry-breaking $F$-terms, which are chosen to be ${\cal
  O}(100\, {\rm GeV})$ thus solving the $\mu$-problem.  
In fact,
the use of singlets is common for solving this problem in models of
direct mediation (see, e.g.~\cite{Giudice:1998bp}). 
As we will discuss, this singlet 
also plays an important role in generating a viable messenger spectrum. 

In the previous section we showed that in the presence of large
singlet couplings, the ISS extremum at $M=0$ may be destabilized. 
It is therefore natural to ask whether the same is true for the
Pentagon model.
While the model is non-calculable, we will argue that for
small quark masses, the singlet coupling must be sufficiently large 
in order to
avoid negative contributions to the MSSM scalars.
Therefore, destabilization is likely to occur.

To see first that non-calculable corrections are crucial in this
setup, let us briefly 
review the model.
The model has just one gauge singlet $S$. 
The relevant part of the superpotential is,
\begin{eqnarray}
  \label{eq:24}
  W= m_Q \tr M + \lambda S \tr YM +\frac{1}{6} g S^{3}+
{\cal A}\left(\det M-\frac{B_+^2}{2}+\frac{B_-^2}{2}-\Lambda^{10}\right)
\end{eqnarray} 
Here $Y_{ij}$ is the hypercharge generator, 
normalized to be $Y_{ij}= {\rm diag}(1,1,1,-3/2,-3/2)$.
As before, it is straightforward to check that near the origin of the
mesonic direction, the potential is extremized at $M_{ij}=0$ for
$i\neq j$.  Furthermore, given the $(SU(3)\times SU(2))_{\rm diag}$
symmetry, the ansatz we are seeking is of the form,
\begin{eqnarray}
  \label{eq:26}
  M_{ij}=M_d\delta_{ij}+M_Y Y_{ij}\,.
\end{eqnarray}
Ignoring first higher order terms in the K\"ahler potential, the
potential is extremized along the baryonic branch at $S=M_Y=0$, with
${\cal A}=0$ and $M_d$ undetermined.

Near the origin, higher order K\"ahler
terms will generate supersymmetry-breaking masses-squared of
order $F_{M_d}^2\sim m_Q^2$ for the mesons, just as in
eqn.~(\ref{corrections}).  They will also shift ${\cal A}$ from zero, so that
${\cal A}\propto F_{M_d}$.  This in turn will generate 
masses from the ${\cal A}\det M$ term in the superpotential,
for both the fermion and scalar mesons. 
Clearly however, as long as the
mesons are close to the origin, these tree-level contributions cannot
dominate over the K\"ahler contributions.  For {\it small} $\lambda$,
the existence of a minimum therefore depends on the 
signs of the parameters $\alpha$, $\beta$ and $\gamma$.

Imagine then that $\lambda$ is small, and that $\alpha$, $\beta$ and
$\gamma$ are such that a minimum is generated.  As we discussed above,
the SM gauge group is embedded in the SU(5)$_{diag}$ flavor symmetry.
Below $\Lambda$, the messengers of gauge mediation are therefore the
mesons.  These get Dirac masses from two sources.  The first is the
${\cal A} \det{M}$ term discussed above, and the second is
higher-dimension K\"ahler terms, such as the third and fourth terms of
eqn~(\ref{kahler}).  Both contributions are proportional to $m_Q$ and
some positive power of $\lambda$.  In addition, the scalar mesons have
supersymmetry-breaking masses of order $m_Q$.  For small $\lambda$,
the messenger supertrace is therefore positive.  Furthermore, for
$\lambda\ll 1$ and $m_Q\ll\Lambda$, there is some region of energies
in which the messengers are weakly coupled.  The positive supertrace
then generates a negative contribution to the masses of MSSM
scalars~\cite{Poppitz:1996xw}.  This contribution arises at one-loop,
and is logarithmically enhanced as $\log(\Lambda_{UV}/m_F)$, where
$\Lambda_{UV}$ is the appropriate cutoff, and $m_F$ is the messenger
scale. In the case at hand, $\Lambda_{UV}\sim\Lambda$, where the
positive supertrace is canceled by additional strongly-interacting
fields charged under the SM gauge group.  Of course close to $\Lambda$
the theory becomes strongly interacting, and there will be
non-calculable corrections to the soft masses.  Nonetheless, for a
large enough scale separation the negative contribution would win
because of the logarithmic enhancement.  We conclude that the coupling
$\lambda$ cannot be too small.  For $\lambda$ of order one, the
minimum would probably be destabilized, much like we found in
section~\ref{singlets}.  In any case, such a large coupling drives the
mesons to VEVs of order $\Lambda^2$.  Thus K\"ahler corrections
are important to all orders, and the minimum required for the Pentagon
no longer depends merely on $\alpha$, $\beta$ and $\gamma$.  It is
therefore not directly related to the ISS conjecture.

Finally, note that we assumed here $m_Q\ll\Lambda$.  In this regime,
the ISS analysis for $\nf > N$ is reliable, and the lifetime of the
minimum is parametrically enhanced.  In~\cite{Banks:2006ma}, $m_Q$ is
taken to be of order $\Lambda$, and the spectrum cannot be reliably
computed.

\section{Conclusions}
\label{sec:conclusions}
ISS conjecture a DSB minimum for $N_f=N$ SQCD.  They reach this
conclusion by deforming the theory with an additional flavor.  The
importance of this conjecture lies in its appeal for model building
and in particular for constructing models of direct mediation which do
not suffer from Landau poles at low energy.  In this paper we
revisited this conjecture.  We argued that deforming the theory gives,
by construction, no information on the existence of such a minimum and
therefore there is no evidence for a DSB vacuum.  In particular, the
existence of this state depends on the signs of three  non-calculable
parameters in the K\"ahler potential.

To demonstrate our point, we studied another deformation by coupling
singlets to the mesons and baryons of the theory.  For sufficiently
large couplings the theory is calculable close to the origin.  As we
showed, the would-be ISS minimum is destabilized by the presence of the
singlets and becomes a saddle point.  Two conclusions are to be
inferred from this deformation: (i) As we dial couplings, a minimum in
one theory becomes a saddle point in another.  This transition occurs
in a region where the theory is non-calculable.  This is in accord
with our claim that no information can be extracted on the existence
of a minimum in the original theory. (ii) Coupling singlets to such
gauge theories can quite generically destabilize existing minima.

Given the latter conclusion we briefly discussed direct mediation
based on $N_f=N$ SQCD, assuming that a minimum does exist.  An example
of such a model is the Pentagon model presented
in~\cite{Banks:2006ma}.  We argued that the coupling to the singlet
cannot be too small in this case.  On the other hand, a large coupling
would drive the mesons far from the origin where both the tree-level
and non-calculable corrections are important.  Thus while likely, one
cannot conclude whether similar destabilization occurs in this model.
Still the existence of the minimum depends on the complete structure
of the K\"ahler potential and is unrelated to the original ISS minimum
at the origin.

\section{Acknowledgments}
We thank Yaron Antebi, Hitoshi Murayama, Yossi Nir, Yaron Oz, Michael
Peskin, Raman Sundrum and Neal Weiner for useful discussions.  We
especially thank Yuri Shirman for many useful discussions, for
collaboration at early stages of this project and for critical reading
of the manuscript.
We also thank Tom Banks and Nathan Seiberg for critical reading of the
manuscript and for many useful remarks.  The work of TV is
partly supported by a grant from the United States-Israel Binational
Science Foundation (BSF), Jerusalem, Israel.  TV would like to thank
the KITP for its hospitality during the course of this project.  The
research of YS and AK is partly supported by the United States-Israel
Science Foundation (BSF) under grant 2002020 and by the Israel Science
Foundation (ISF) under grant 29/03. The research of AK is also
supported by the Gutwirth Foundation.

\appendix
\section{The ISS-like extremum: full analysis}
\label{low}
In~\ref{sec:non-supersymm-saddle} we presented a somewhat simplified
analysis of the $\nf=\nc$ theory coupled to singlets, keeping the
Lagrange multiplier in the theory and treating it on equal footing
with the other fields.  Here we will refine this analysis, and impose
the constraint right away to eliminate the heavy field $B_-$, whose
mass is of order $\Lambda$.  For convenience, we will set $\Lambda=1$,
such that all fields are dimensionless.  The quantum modified
constraint then gives
\begin{eqnarray}
  \label{bdef}
  B_{-}=\sqrt{2-2M^{N}+B_{+}^{2}}\ .
\end{eqnarray}
Using the parametrization~\eqref{diagansatz} 
the superpotential is then 
\begin{eqnarray}
  \label{eq:3}
  W_{eff}=N \lambda  M S+\kappa T_{+}B_{+}+\kappa T_{-}\sqrt{2-2M^{N}
    +B_{+}^{2}}+N m_Q M+
  \frac{N}{2}m_S S^{2}+\frac{m_{T}}{2}\left(T_{+}^{2}+T_{-}^{2}\right),
\end{eqnarray}
The potential is extremized for
\begin{eqnarray}
  \label{v1}
  0= \frac{\partial V}{\partial S} &=&  N m_S F_{S}^{*}+
  N\lambda F_{M}^{*}
  \\
  \label{v2}
  0=\frac{\partial V}{\partial M}  &=&  N \lambda F_{S}^{*}+
  \frac{N\kappa T_-B_{+}M^{N-1}}{B_{-}^{3}}F_{B_{+}}^{*}-
  \frac{N\kappa M^{N-1}}{B_{-}}F_{T_{-}}^{*}-
  \\ \nonumber
  &&\frac{N(N-1)\kappa T_{-}M^{N-2}}{B_{-}}F_{M}^{*}
  -\frac{N^{2}\kappa T_{-}M^{2N-2}}{B_{-}^{3}}F_{M}^{*}
  \\
  \label{v3}
  0=\frac{\partial V}{\partial T_{-}} & = & m_{T}F_{T_{-}}^{*}-
  \frac{N\kappa M^{N-1}}{B_{-}}F_{M}^{*}+
  \frac{\kappa B_{+}}{B_{-}}F_{B_+}^*
  \\
  \label{v4}
  0=\frac{\partial V}{\partial T_{+}} & = & m_{T}F_{T_{+}}^{*}+
  \kappa F_{B_{+}}^{*}
  \\
  \label{v5}
  0=\frac{\partial V}{\partial B_{+}} & 
  = & \frac{\kappa T_{-}}{B_{-}}F_{B_{+}}^{*}
  -\frac{\kappa T_-B_{+}^{2}}{B_{-}^{3}}F_{B_+}^{*}
  +\kappa F_{T_{+}}^{*}+
  \\ \nonumber 
  &&\frac{\kappa B_{+}}{B_{-}}F_{T_{-}}^{*}
  +\frac{N\kappa T_{-}B_{+}M^{N-1}}{B_{-}^{3}}F_{M}^{*}
\end{eqnarray}
where we use $B_-$ to denote the combination~(\ref{bdef}) for
convenience.  The last two equations hold if we choose
\begin{eqnarray}
  \label{eq:2}
  F_{T_+}= F_{B+} =0\ ,
\end{eqnarray}
and therefore
\begin{eqnarray}
  \label{eq:4}
  B_+=T_+=0\ .
\end{eqnarray}
Thus, $B_+$ and $T_+$ remain as in~(\ref{ourextr}), and their
$F$-terms still vanish.  At tree level, there are therefore no mass
terms that mix the $(B_+,T_+)$ sector with the $(M,S)$ sector, just as
we found in section~\ref{sec:non-supersymm-saddle} However, a $T_--M$
mixing is generated now.

Since it is difficult to solve the remaining equations exactly, we
will study the theory in the decoupling limit, as an expansion for
small $\lambda$.  It is convenient to choose $m_S$ and $m_T$ of order
one.  In view of the discussion in section~\ref{singlets}, we want
$\lambda$ to be smaller than $\kappa$.  To maintain calculability we
also choose $m_Q\sim \lambda^2$.  We first note that $F_T$ shifts from
zero  since otherwise  eqn~\eqref{v3} isn't satisfied.  To solve
this equation we take the ansatz
\begin{equation}
T_{-}=-\frac{\kappa}{m_{T}}B_- +\frac{\delta
 T_{-}}{m_{T}^{2}}\label{eq:tplus}
\end{equation}
which gives, 
\begin{eqnarray}
  \label{eq:6}
  m_TF_{T_-}^* = \delta T_-^* = \frac{N\kappa M^{N-1}}{B_-}F_{M}^{*}.
\end{eqnarray}
Furthermore, in the decoupling limit $M$ is small 
(we will see below that it is of
order $\lambda^{\frac{2}{N-2}}$), so we can safely neglect terms of
order $\left(M^{2N}\right)$~\footnote{For convenience we work here 
in a limit of
  large $N$.}. Using the above, we find to leading order,
\begin{eqnarray}
  \label{map}
  M\simeq \left(\frac{\lambda^{2}}{m_S}\frac{m_{T}}{\kappa^{2}}
    \frac{1}{(N-1)}\right)^{\frac{1}{N-2}}
\end{eqnarray}
just as in eqn~(\ref{ourextrms}), and 
\begin{eqnarray}
   \label{sap}
   S &=& -\frac{\lambda^*}{|m_S|^2+|\lambda|^2}\left(m_Q +
     \frac{\lambda}{\lambda^*}m_S^* M - \frac{\kappa T_- M^{N-1}}{B_-}\right)
   \simeq -\frac{\lambda^* m_Q}{|m_S|^2}-\frac{\lambda^* M}{m_S}
 \end{eqnarray}
It is easy to see that this coincides with  the
solution~(\ref{ourextrms}). 

Having found the extremum, we can now calculate the bosonic
mass-squared matrices.  As we mentioned above, at tree-level, there is
no mixing between $B_+$, $T_+$ and the remaining fields.
The $(B_+,T_+)$ mass matrix is
\begin{eqnarray}
  \label{eq:7}
  m_{BT}^{2}=\left(\begin{array}{cccc}
      |\kappa|^{2}+\frac{|\kappa|^{4}}{|m_{T}|^{2}} & 
      m_{T}\kappa^* -\frac{\kappa^{*2}\kappa}{m_{T}^*} & 
      \xi_{BT} &0\\
      m_{T}^*\kappa-\frac{\kappa^2\kappa^*}{m_{T}} & |\kappa|^{2}+|m_{T}|^{2} 
      & 0 & 0\\
      \xi_{BT}^*
      & 0 & |\kappa|^{2}+\frac{|\kappa|^{4}}{|m_{T}|^{2}} & 
      m_{T}^*\kappa-\frac{\kappa^2\kappa^*}{m_{T}}\\
      0 & 0 & m_{T}\kappa^* -\frac{\kappa^{*2}\kappa}{m_{T}^*} & |\kappa|^{2}+|m_{T}|^{2}
    \end{array}\right)
\end{eqnarray}
where 
\begin{eqnarray}
  \label{eq:8}
  \xi_{BT}=\delta T_-\frac{N\kappa  M^{N-1}}{m_T^2 B_-^3}F_M^* \ll \xisb
\end{eqnarray}
As usual the diagonal blocks of this matrix are the mass matrices for
the fermions.  The off-diagonal terms involve the supersymmetry
breaking $F$-term, and are parametrically small. In fact, they are
smaller than the supersymmetry-breaking contributions in the $M,S$
sector, $\xisb$.  Note that this remains true when non-calculable
contributions of the form \eqref{kahler} are taken into account.  The
latter induce susy-breaking masses for $B_+$ which are of order $m_Q$
and therefore larger than $\xi_{BT}$ but smaller than $\xisb$.  
Clearly, all eigenvalues of
this matrix are positive and no instability develops here.

We now turn to the second sector which contains the fields 
$M$, $S$ and $T_-$. Calculating the
boson matrix, substituting~(\ref{map},\ref{sap}) for 
the VEVs and neglecting terms of order $M^{2N}$ or higher we get
{\scriptsize 
\beq
  \label{eq:10}
  m_{0}^{2}=N^{2}\left(\begin{array}{cccccc}
      |\lambda|^2+ \left|\frac{\lambda^{2}}{m_S}\right|^2+|\Omega|^{2} & 
      m_S^*\lambda+\frac{\lambda^*\lambda^2}{m_S} & \Omega^*\frac{\lambda^{2}}
{m_S}+\frac{m_T^*}{N}\Omega       & \xisb & 0 & m_{MT_-}^2 
      \\
      m_S\lambda^*+\frac{\lambda^{*2}\lambda}{m_S^*} & |m|^{2}
+|\lambda|^{2} & \Omega^*\lambda      & 0 & 0 & 0
      \\
      \Omega\frac{\lambda^{*2}}{m_S^*}+\frac{m_T}{N}\Omega^*  
& \Omega\lambda^* & \frac{|m_{T}|^{2}}{N^{2}}+|\Omega|^{2} & m_{MT_-}^2 & 0 & 0
      \\
      \xisb & 0 & m_{MT_-}^{*2}& |\lambda|^2+ 
\left|\frac{\lambda^{2}}{m_S}\right|^2+|\Omega|^{2} & 
      m_S\lambda^*+\frac{\lambda\lambda^{*2}}{m_S^*} & 
\Omega\frac{\lambda^{*2}}{m_S^*}+\frac{m_T}{N}\Omega^* 
      \\
      0 & 0 & 0 & m_S^*\lambda+\frac{\lambda^2\lambda^*}{m_S} & 
|m|^{2}+|\lambda|^{2} & \Omega \lambda^*
      \\
      m_{MT_-}^{*2}& 0 & 0 &
      \Omega^*\frac{\lambda^{2}}{m_S}+\frac{m_T^*}{N}\Omega  
& \Omega^*\lambda & \frac{|m_{T}|^{2}}{N^{2}}+|\Omega|^{2} 
    \end{array}\right) \nonumber
\eeq
}
where $\xisb$ is as in \eqref{eq:1} and we  defined  
\begin{eqnarray}
  \label{eq:12}
  \Omega&\equiv& \frac{1}{N}\frac{\partial F_{M}}{\partial T_{-}}=
\frac{1}{N}\frac{\partial F_{T_{-}}}{\partial M}=
  -\frac{\kappa M^{N-1}}{B_-}\ ,
  \\
  m_{MT_-}^2& \equiv & \frac{\kappa m_Q^* (N-1)M^{N-2}}{\sqrt{2}}~.
\end{eqnarray}
Since we are looking for supersymmetry-{\it breaking} effects, 
we should carry out our calculations up to the order of the largest
non-vanishing contribution in the off-diagonal block, namely
$m_{MM}\sim {\cal O}(\lambda^4)$.  Fortunately, the situation is
simplified by noting that this sector includes the Goldstino.  Indeed the
$3\times3$ blocks on the diagonal coincide with the fermionic mass
matrix and hence vanish.  This is sufficient to show  that the determinant 
of this matrix is negative without keeping track of such small
orders in $\lambda$.  We thus get the same instability as 
we had in section~\ref{singlets}. 

It is worth noting that any contribution from the CW potential
would be proportional to either $\xisb$ or to $\xi_{BT}$.
On the other hand, the tree-level tachyonic mass is proportional
to $\xisb$. Since $\xi_{BT}<\xisb$, the CW contribution cannot
compete with the tree-level contribution.



\end{document}